\documentclass[prl,twocolumn,a4paper,amsmath,floatfix]{revtex4}
\usepackage[utf8]{inputenc}
\usepackage{graphicx}
\usepackage{amsmath}

\usepackage{color}
\usepackage[utf8]{inputenc}
\usepackage{graphicx}
\usepackage{amsmath}

\begin{document}
\title{Liquids more stable than crystals}
\author{Frank Smallenburg and Francesco Sciortino}

\affiliation{ Department of Physics, Sapienza, Universit$\acute{a}$ di Roma, Piazzale Aldo Moro 2, I-00185, Roma, Italy }
\maketitle
%
%
{\bf 
All liquids (except helium due to quantum effects) crystallize at low temperatures, forming ordered structures.
The competition between disorder, which stabilizes the liquid phase, and energy, which favors the ordered crystalline structure, inevitably turns in favor of the latter when temperature is lowered and the entropic contribution to the
free energy becomes progressively less and less relevant.  The ``liquid'' state survives at low temperatures only
as a glass, an out-of-equilibrium arrested state of matter.  This textbook description holds inevitably for
atomic and molecular systems, where the interaction between particles is  set by quantum mechanical laws.   The question remains whether the same physics hold for colloidal particles, where inter-particle interactions are usually short-ranged and tunable.   Here we show that for patchy colloids with  limited valence, conditions can be found for which the  disordered liquid phase is stable all the way down to the zero-temperature limit.  Our results offer interesting cues for understanding
the stability of gels and the glass forming ability of atomic and molecular network glasses. 
}

The ability to control --- via chemical\cite{granickpatchy2011,pinenature,ilona}  or physical\cite{kraft} patterning --- the selectivity and angular flexibility  of the inter-particle interaction\cite{glotzer} makes it possible to provide valence to colloids.   Exploiting valence  offers enormous possibilities, some  of which have been addressed in recent years, theoretically \cite{emptyliquids}  numerically\cite{bicontinuouspatchy,equilibriumgels3,frenkellowtempfluid1}   and experimentally\cite{pinenature,laponite,dnabellini}.   Since colloidal interactions are typically short-ranged
and thus limited to  nearest neighbors, valence provides an implicit quantization of the particle energy, which becomes essentially proportional to the (limited) number of  formed bonds.  
In an optimal arrangement, all particles  bind with $f$ neighbors and the system is
in its lowest possible (ground) energy state.  Typically, such an optimal arrangement is spatially ordered, defining the most stable crystal phase(s).

In principle, when density is not too high, it is possible to envision disordered optimal arrangements, in which all   particles form exactly $f$ bonds,  giving rise to a disordered fully bonded structure which has exactly the same energy as the crystal.    Under this unconventional condition,
 the stability of the system becomes controlled by its entropy, a case reminiscent of   hard-sphere (HS) colloids, where  crystal formation is observed despite the absence of  any cohesive energy.  
 In this study we demonstrate that the flexibility of the bond (encoded in the angular
 patch width) --- a tunable quantity in the design of patchy colloidal particles ---  is the key element in controlling the  entropy of the liquid as compared to the one of the crystal. Large binding angles combined with limited valence give rise to thermodynamically stable, fully bonded liquids.  
 
The model we consider is an extension of the widely used Kern-Frenkel model\cite{KF}  for patchy particles,  where we explicitly enforce a single-bond-per-patch condition. Each spherical particle of diameter $\sigma$ has $f$ circular attractive patches on its surface, characterized by an opening angle $\theta_m$. Two particles are bonded with a fixed bonding energy $\epsilon$ when their center-to-center distance is smaller than the interaction range $\sigma + \delta$, and the vector connecting the particles passes through a patch on both particles (see Supplementary Information (SI)).
We examine a mono-disperse systems of particles with $f=4$ attractive patches, arranged in a tetrahedral configuration, mimicking recently synthesized colloids\cite{pinenature}. 
The values of $\delta$ and $\theta_m$ fix the volume available for bonding $V_b$.

\begin{figure*}
\begin{center}
\begin{tabular}{cc}
  \begin{tabular}{cc}
  \includegraphics[width=0.32\textwidth]{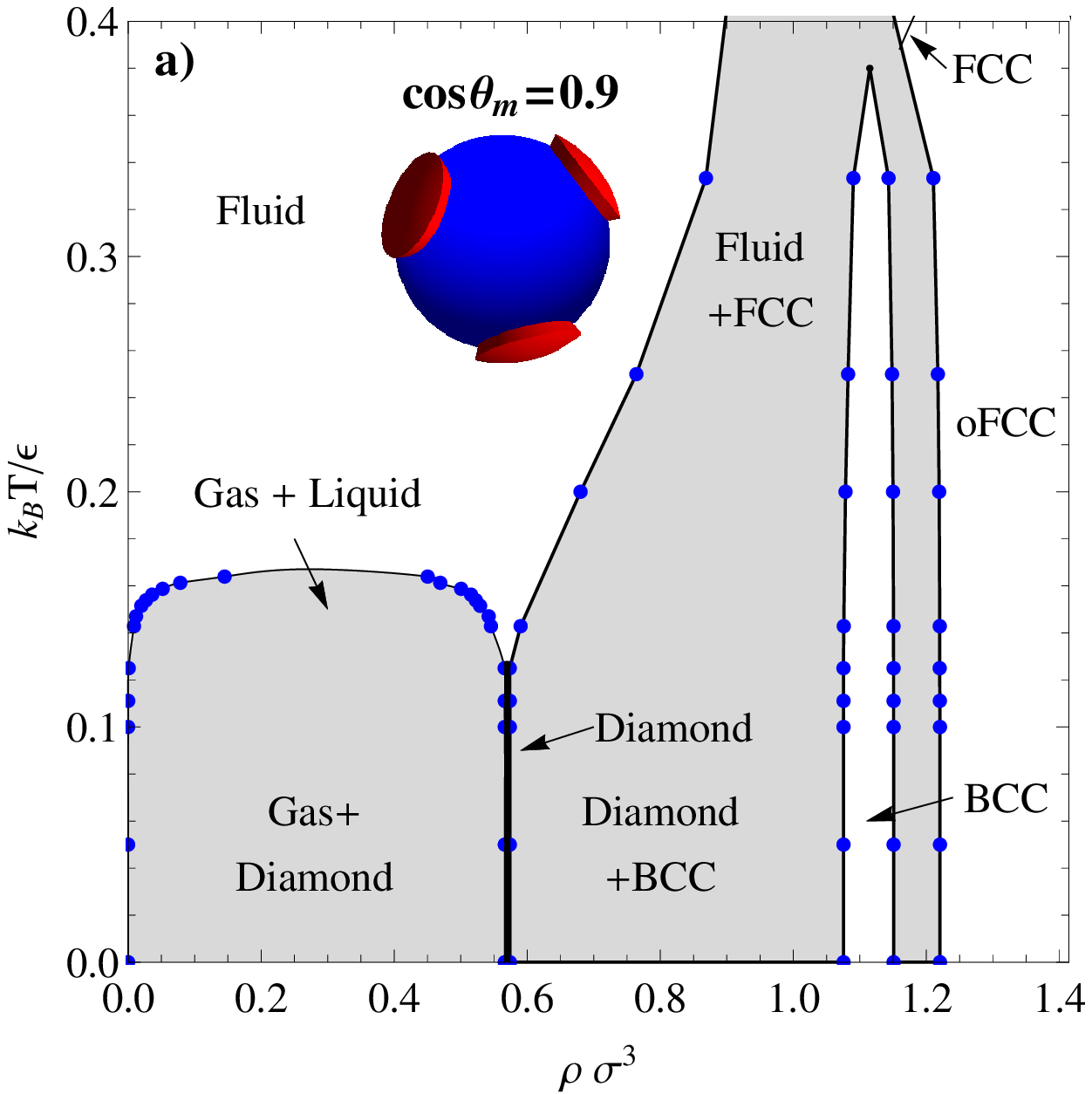} & \includegraphics[width=0.32\textwidth]{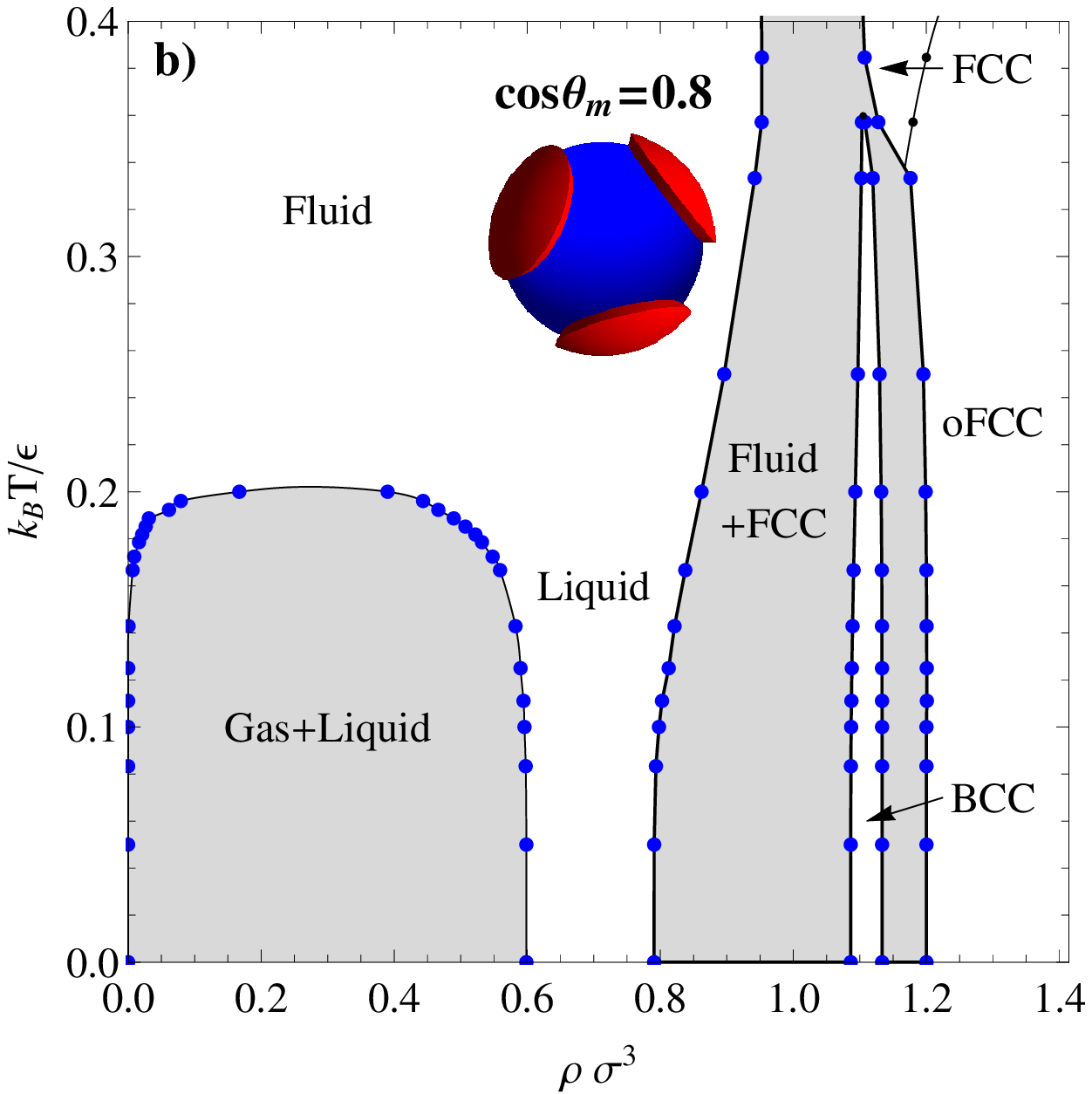} \\  
  \includegraphics[width=0.32\textwidth]{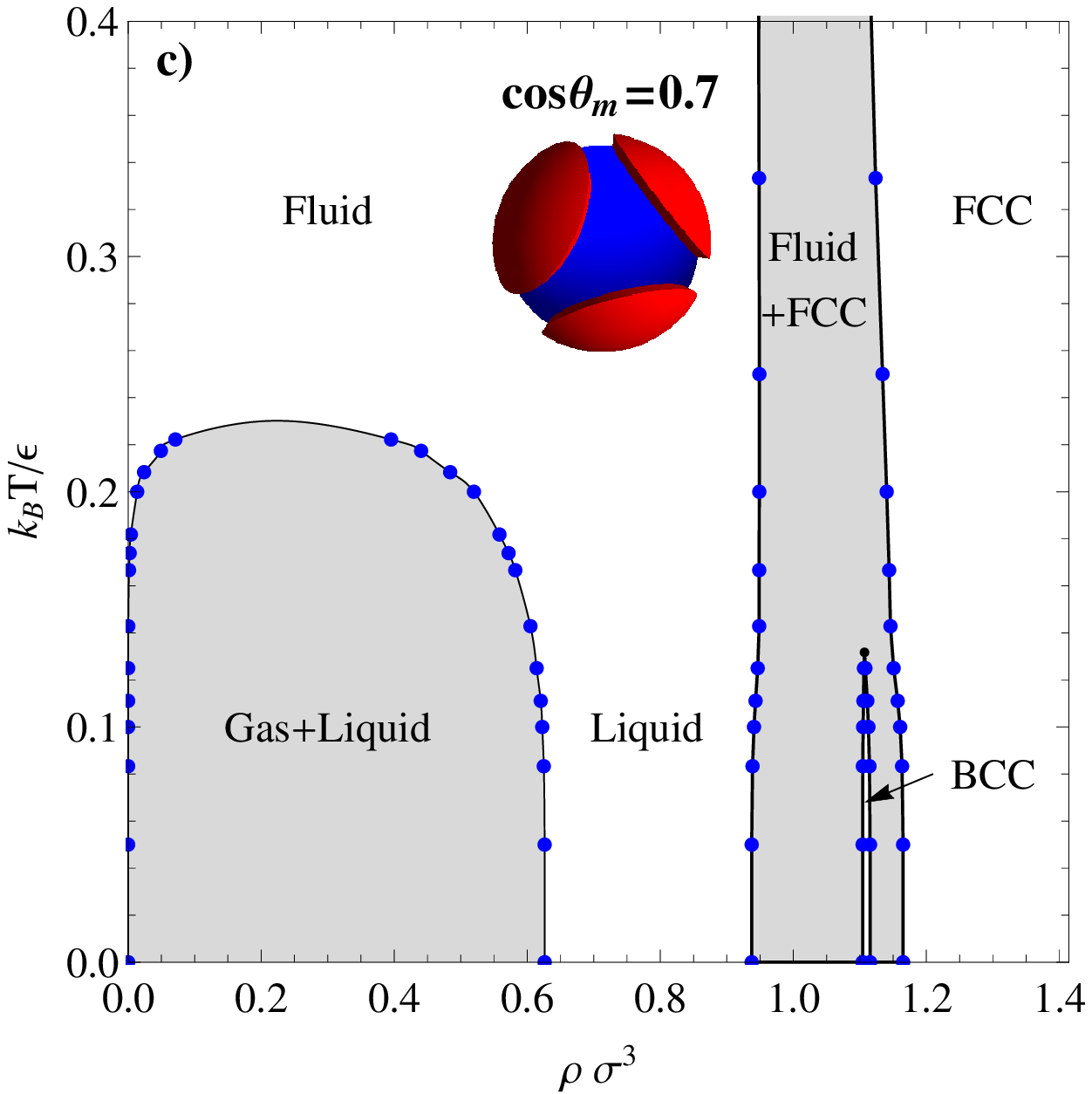} &\includegraphics[width=0.32\textwidth]{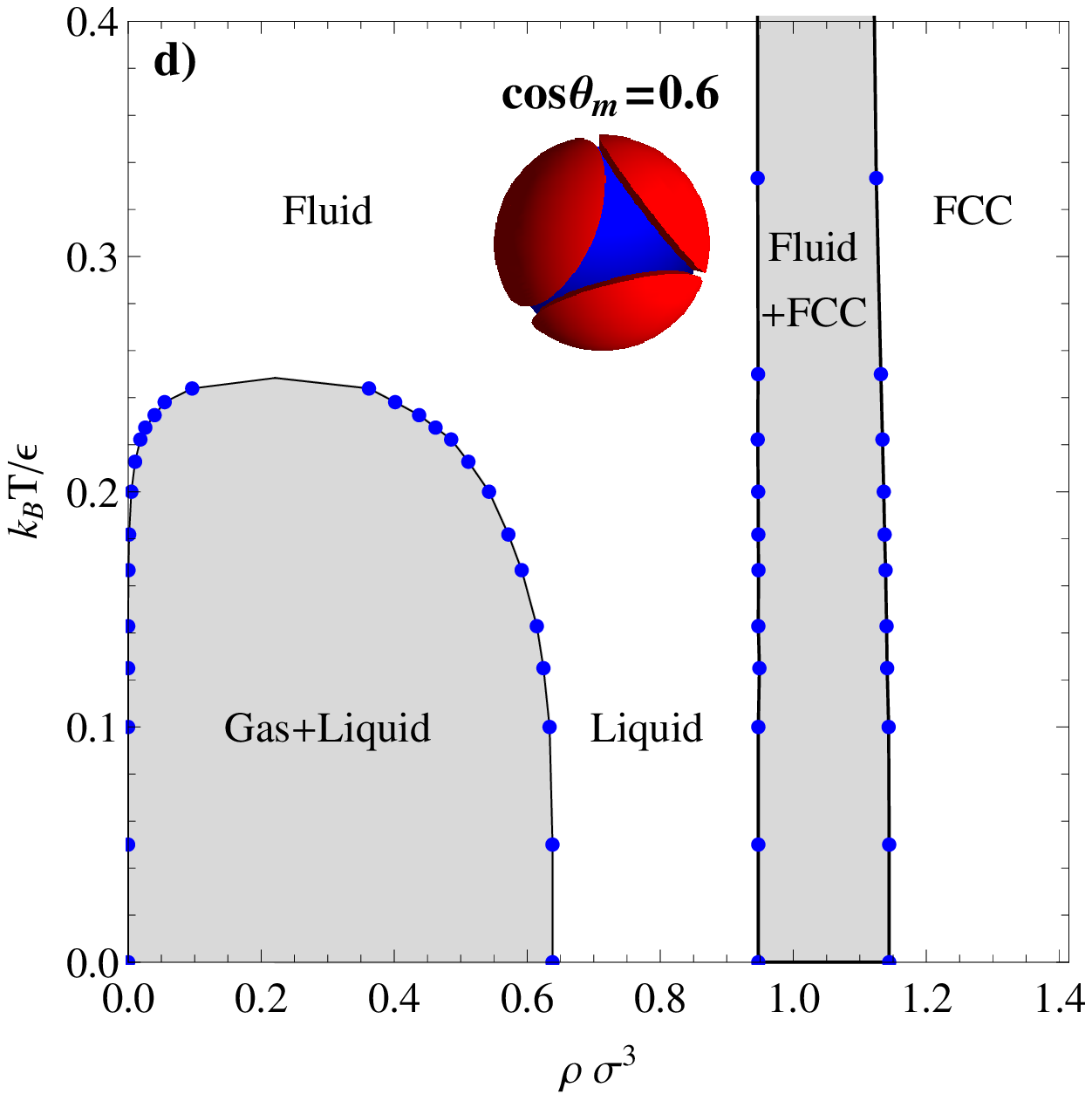}  
  \end{tabular} & 
  \begin{tabular}{l}
  {\bf e)} \\
   \includegraphics[width=0.14\textwidth]{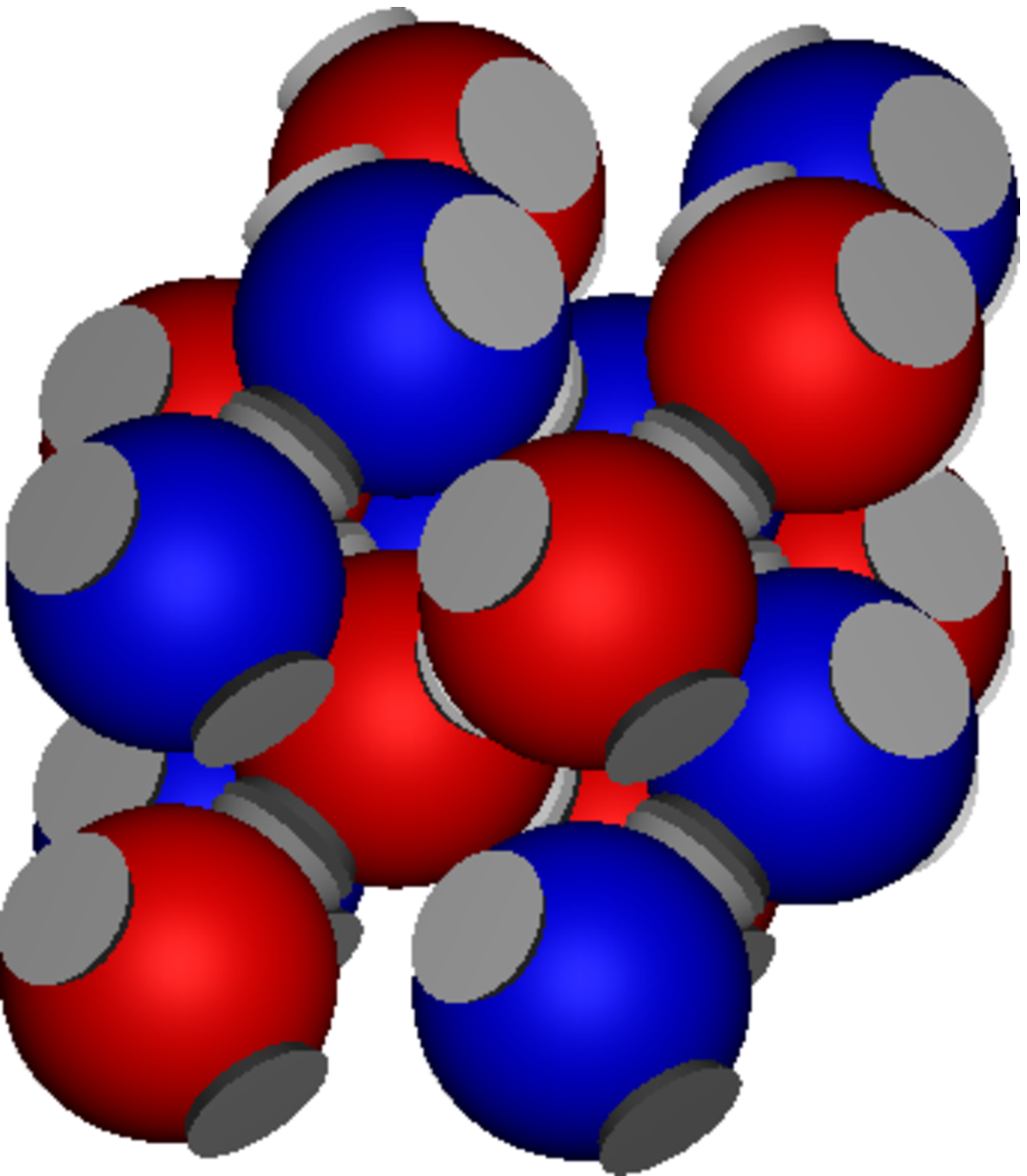}    \\
   \multicolumn{1}{c}{BCC}\\
  \\
  {\bf f)} \\
  \includegraphics[width=0.14\textwidth]{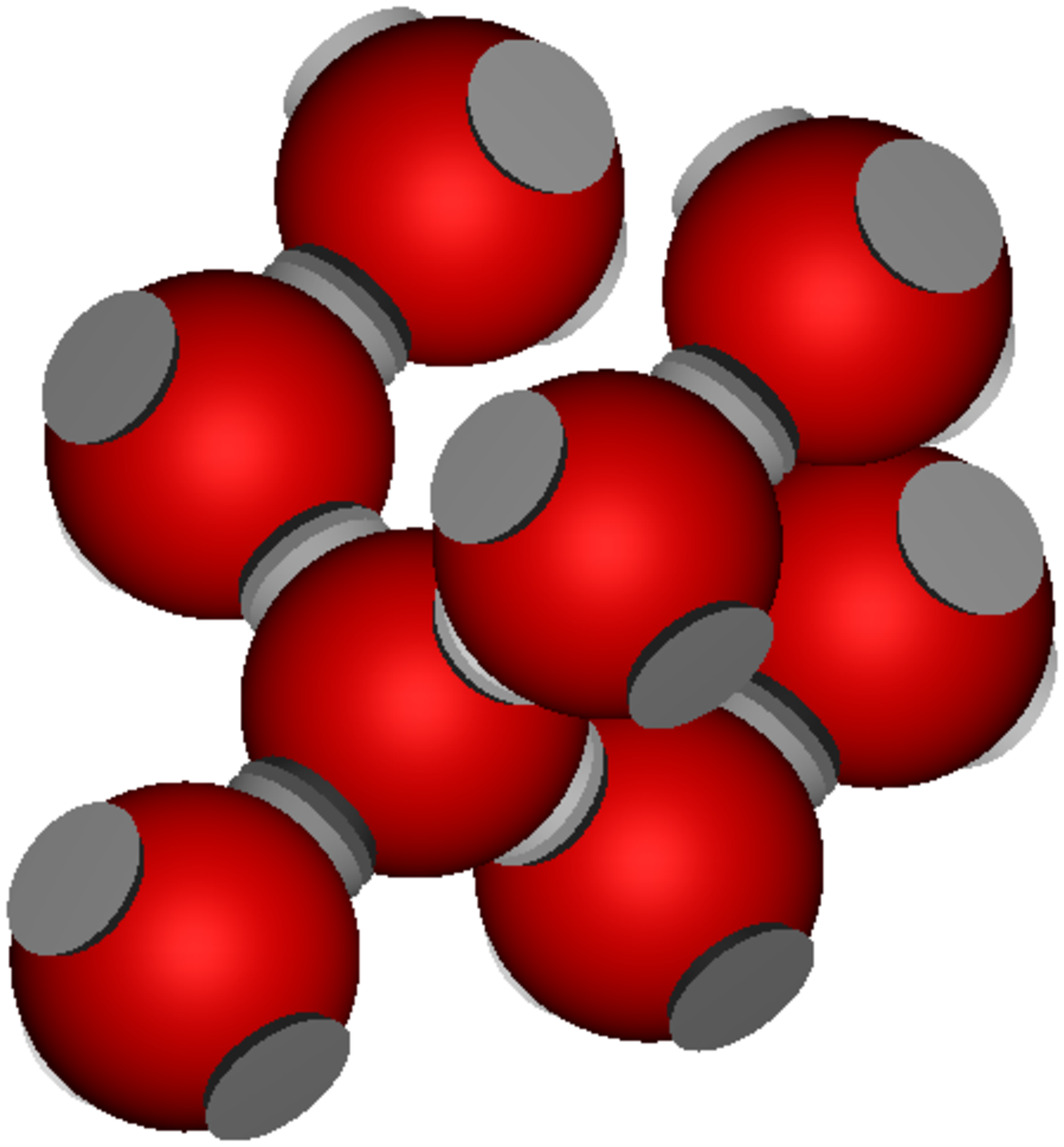}    \\
   \multicolumn{1}{c}{Diamond}\\
  \\
  {\bf g)} \\
  \includegraphics[width=0.14\textwidth]{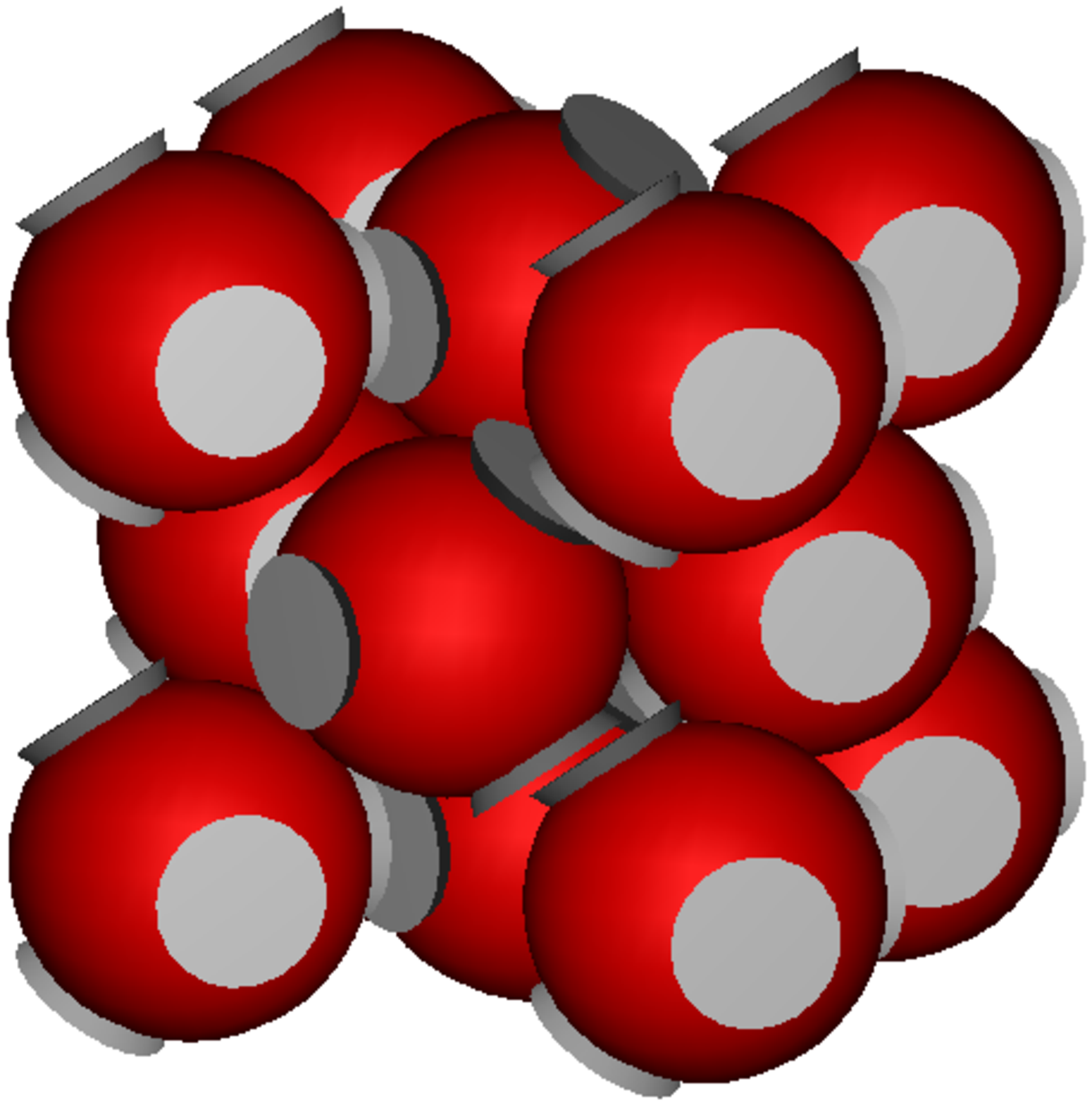}    \\
   \multicolumn{1}{c}{FCC}\\
  \end{tabular}
\end{tabular}
\caption{ {\bf Phase diagram of  tetrahedral coordinated patchy colloids   for different  patch widths.}
(a)  $\cos \theta_m = 0.9$ , (b) $\cos \theta_m =0.8$, (c) $\cos \theta_m =0.7$ , and (d) $\cos \theta_m =0.6$.  Data are reported   in  dimensionless density ($\rho \sigma^3$) and temperature  $k_BT/\epsilon$, where $k_B$ is the  Boltzmann constant. The interaction range in all cases is fixed at $\delta = 0.12 \sigma$, a value typical of colloidal interactions. Narrower patches ($\cos \theta_m >0.9$) behave as the $\cos \theta_m = 0.9$ case\cite{kernfrenkelphasediags} and are not shown here. 
The gray areas denote coexistence regions and the points denote calculated coexisting states (tie lines are horizontal). The points below $k_B T/\epsilon = 0.1$ are based on extrapolation of the potential energy. The label oFCC denotes the orientationally ordered face-centered cubic phase, while the disordered FCC phase is simply denoted as FCC. The pictures on the right  show the unit cell of BCC (e), with the two interspersed diamond lattices indicated in different colors, diamond (f)
and orientationally ordered FCC (g). 
} \label{phasediagrams}
\end{center}
\end{figure*}

 The phase diagrams, evaluated using free-energy calculations, for different $\theta_m$ are shown in Fig.~\ref{phasediagrams} as a function of  density $\rho$  and  $T$. 
The phases appearing in Fig.~\ref{phasediagrams} include, besides the disordered gas and liquid phases, a diamond cubic crystal phase, a body-centered cubic (BCC) crystal phase, consisting of two interlocking diamond lattices, and a face-centered cubic (FCC) crystal phase, where, for sufficiently narrow patches, the bonds are periodically ordered at low $T$ and disordered at high $T$.  All these crystals are  fully bonded at low $T$.

For  narrow patch width,  the phase diagram closely resembles the standard phase diagrams with a triple point below which the liquid state ceases to exist. Unexpectedly, for  wider patches the open diamond crystal phase disappears from the phase diagram and a region in which the liquid is stable down to  vanishing $T$  opens up at intermediate $\rho$, in agreement with a numerical study of  DNA-coated colloids\cite{frenkellowtempfluid1,frenkellowtempfluid2}.  In this range of 
densities, the  liquid is   the phase with the lowest free energy, despite its intrinsic long range disorder.  
Increasing the patch width even further reduces the region of stability for the BCC and the disordered FCC crystals.  For very wide patches, the phase diagram simply consists of  gas, liquid, and FCC crystal phases, with a large stable liquid region even in the zero-temperature limit.

\begin{figure}
\begin{center}
\hspace{0.014\textwidth}\includegraphics[width=0.436\textwidth]{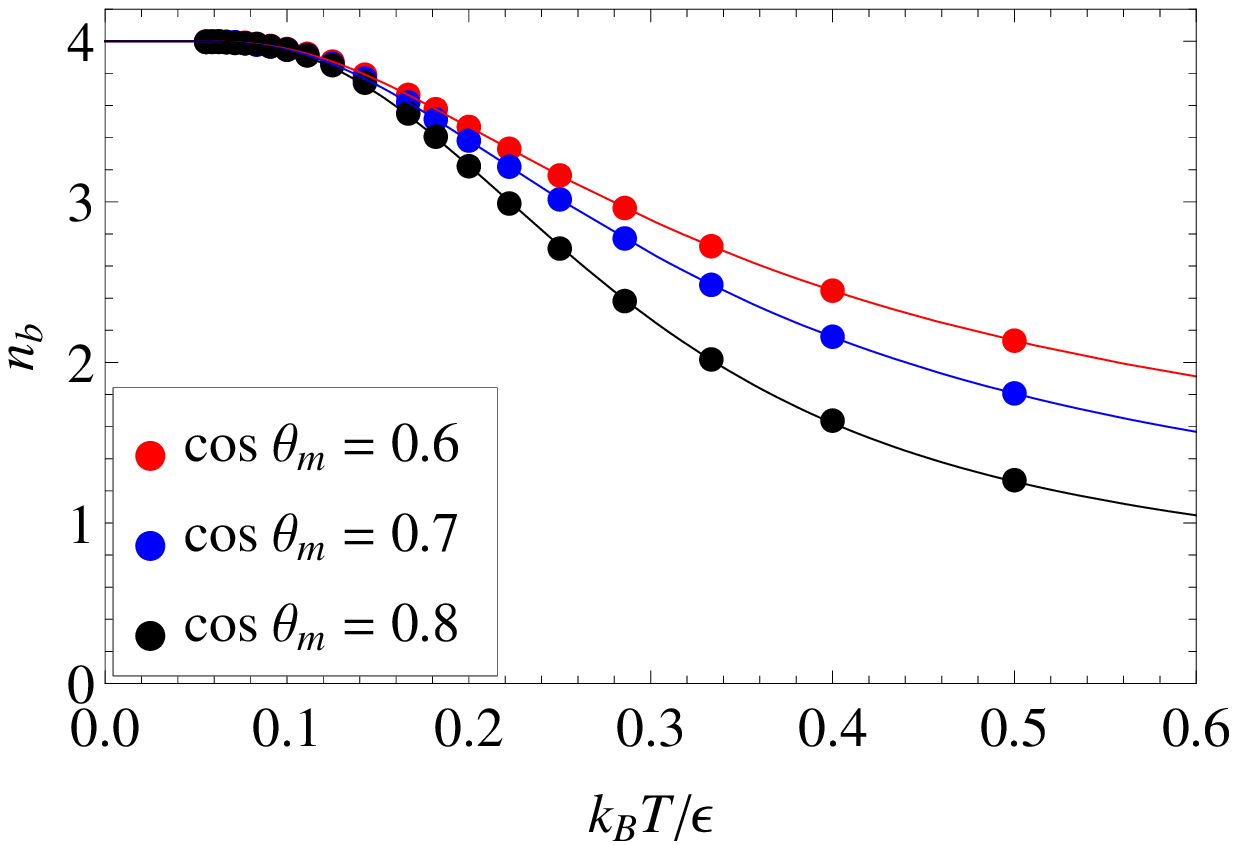}
\end{center}
\caption{ {\bf Number of bonds per particle  in the liquid phase}.
Here $\rho \sigma^3 = 0.57$, the same   density of the diamond crystal phase. In the zero-temperature limit, the number of bonds per particle $n_b$  reaches the maximum of $4$. Points denote measurements in $NVT$ simulations, and  lines are fits to guide the eye. Note that the approach to the ground state ($n_b=4$)  takes place at higher $T$ for large $\theta_m$  values,  a result induced by the larger  bonding volume. 
}
\label{fig:nb}
\end{figure}

Why is  the liquid  more stable than the crystal?  Figure \ref{fig:nb} shows the $T$ dependence of the number of bonds per particle $n_b$  for different $\theta_m$.  Upon cooling, $n_b$ increases continuously to four,  while the system progressively approaches the fully bonded random tetrahedral network state, the ground state of the system.  Configurations with $n_b=4$ are indeed sampled during the numerical runs. At low $T$,  the potential energy of the liquid  is thus
equal to that of the crystals. 
 Hence, the stability of the liquid, 
as in the HS case,  results from  a subtle competition between vibrational $S_\mathrm{vib}$  and configurational   $S_\mathrm{conf}$  components of the total  entropy $S_\mathrm{tot}= S_\mathrm{vib} + S_\mathrm{conf}$ \cite{moreno,debenedetti}.  $S_\mathrm{vib}$  measures  the (phase-space) volume explored by each particle in a fixed bonding topology,  and is larger in the ordered  structure than in the  fluid. $S_\mathrm{conf}$  measures the number of distinct configurations resulting in a  fully bonded macroscopic state. It vanishes in the crystals and is positive in the fluid phase.  Both quantities can be calculated as discussed in the methods section.

\begin{figure}
\begin{center}
\includegraphics[width=0.45\textwidth]{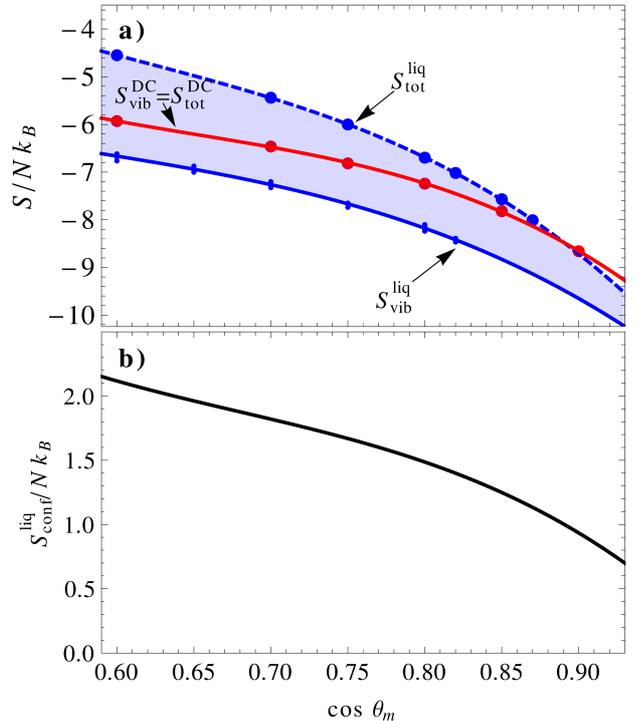}
\end{center}
\caption{ {\bf  Total, vibrational, and configurational entropy for the zero-temperature  phases}.
Here  $\rho \sigma^3 = 0.57$.   {\bf a)} The vibrational and total entropy for the diamond crystal (DC) and liquid (liq) phases. For the diamond crystal, the 
  total entropy $S_\mathrm{tot}^\mathrm{DC}$ coincides with the vibrational entropy
  $S_\mathrm{vib}^\mathrm{DC}$.  Below $\cos \theta_m \approx 0.89$, the fully bonded liquid becomes more stable  than the crystal.  
   For each   angular patch width $ \theta_m$,  five independent $S_\mathrm{vib}^\mathrm{liq}$ calculations are shown.  Lines are polynomial fits to the points. With extremely lengthy simulations  fully bonded configurations can be sampled up to $\cos \theta_m=0.92$.    Note that the entropies are negative since we are investigating a classical system and the (constant) kinetic contribution is not included. 
  {\bf b)} Configurational entropy  $S_\mathrm{liq}^\mathrm{conf}$ of the liquid at $T \rightarrow 0$, calculated as the difference between the fits for  $S_\mathrm{tot}^\mathrm{liq}$ and  $S_\mathrm{vib}^\mathrm{liq}$.
}

\label{entropies}
\end{figure}
 
 Figure ~\ref{entropies}a shows how the patch width affects the entropy.   Similar to dense HS, 
 $S_\mathrm{vib}$ in the diamond structure  ($S_\mathrm{vib}^{DC}$) is always larger than in the liquid ($S_\mathrm{vib}^{liq}$).  Hence
  in the disordered phase each particle is more constrained than in the ordered one. If $S_\mathrm{vib} $ would be the only source of entropy,  the diamond would be the stable
 phase at all $\theta_m$.  However, the entropy of the liquid  is enhanced by $S_\mathrm{conf}$, causing for wide angles the unconventional stability of the liquid phase even when $T \rightarrow 0$.  
 In this large $\theta_m$ region, the model  provides a neat example of a system for which the Kauzmann temperature ($T_K$)\cite{debenedetti}  does not exist. Indeed, $S_\mathrm{conf}$  (Fig.~\ref{entropies}b) does not vanish when $T \rightarrow 0$.  For small $\theta_m$, extrapolations suggest that a finite $T_K$ could exist, but its detection is preempted by dynamic arrest and/or crystallization (as commonly found in atomic or molecular systems). Thus, for wide patches, a previously unexplored thermodynamically stable state of matter arises: the disordered fully bonded network.  We note that while our results  are for $f=4$, it is reasonable to expect that zero-temperature liquids will also occur in systems with a different (but small) $f$, assuming a large patch angle.  Our results provide a clear framework for understanding the microscopic origins of the observed difficulties of crystallizing  DNA-coated colloids when valence is small\cite{frenkellowtempfluid1,frenkellowtempfluid2}. We also stress 
 the relevance of our predictions for the colloidal realm \cite{granickpatchy2011,pinenature,frenkellowtempfluid1,frenkellowtempfluid2}  where bonding rigidity is largely dominated by entropic effects and
    bonds do not become less flexible upon decreasing $T$. This fact is  crucial in maintaining a finite $S_\mathrm{conf}$ in the low-$T$ limit  and explains why the fully bonded network state does not occur in  atomic or molecular systems, where  $\theta_m$ decreases on cooling.

\begin{figure}
\begin{center}
\includegraphics[width=0.45\textwidth]{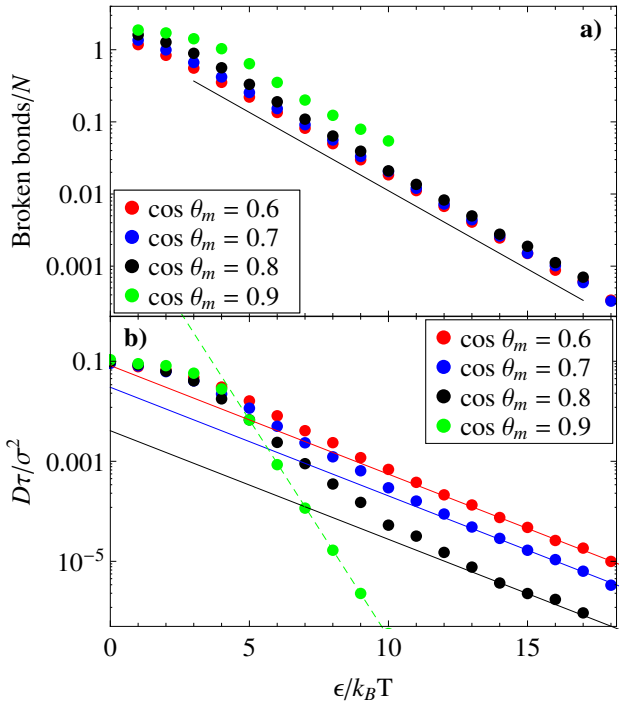} \\
\caption{{\bf Defects of the fully bonded network and their dynamic role.} {\bf a)} The number of broken bonds per particle $\alpha_{bb}$ as a function of the inverse temperature $\epsilon/k_B T$, for different patch widths, at constant density $\rho \sigma^3 = 0.65$. The line has slope $-1/2$.
{\bf b)} The (a-dimensional) diffusion coefficient $D$ as a function of the inverse temperature $\epsilon/k_B T$.  In the EDMD, an attempt to break, form or switch bond is performed with a rate $\gamma$. Data here refer to a constant bond switching rate $\gamma \tau = 100$, a value for which the dynamics is not affected by the specific value of $\gamma$  (see SI).
The lines have slopes $-1/2$ (for $\cos \theta_m = 0.6, 0.7, 0.8$) and slope $-2$ (for $\cos \theta_m = 0.9$).} \label{diffusionplots}
\end{center}
\end{figure}
    
To provide further evidence that sampling of the  fully bonded state is not pre-empted by  dynamic arrest and to characterize the low $T$ dynamics we investigate the microscopic mechanism
associated with the evolution of the network. 
We start by focusing on the defects of the fully bonded network (broken bonds), which act as elementary diffusing units.
At any $T$ close to zero the number of broken bonds per particle ($\alpha_{bb}$)   depends on $T$  as  $\exp(-\epsilon/2k_BT)$,  as demonstrated in the SI. Fig.~\ref{diffusionplots}a shows that $\alpha_{bb}$  indeed satisfies the  exponential $T$ dependence
for all $\theta_m$.  We observe two 
different mechanisms for restructuring the network at low $T$:  bond-breaking and bond-switching,
depending on $\theta_m$.   For small
$\theta_m$  the single-bond-per-patch condition is automatically implemented 
by geometric constraints\cite{emptyliquids} and hence bonds must first break and then 
reform with new (or identical) partners.  For large
$ \theta_m$ the bonding volumes of different neighbors can overlap and the
switching of bonds with no energetic penalty allows the system to relax stresses relatively quickly. The bond-switching mechanism mimics the  microscopic dynamics  of 
vitrimers\cite{vitrimers}, a recently invented malleable network plastic  where a catalyst enables the switching (transesterification) of bonds between nearby polymers.

To quantify the low $T$ dynamics  we perform event driven molecular dynamics (EDMD) simulations and  examine the diffusion coefficient $D$ at low $T$. Results are reported in Fig.~\ref{diffusionplots}b.   For wide patches, $D$ follows an Arrhenius law, with an activation energy of $\epsilon/2$, the same as $\alpha_{bb}$. $D$ is thus  proportional to the number of broken  bonds, demonstrating that 
switching between broken and formed bonds is the key element in the microscopic dynamics.
For patch widths  such that bond switching is not a viable mechanism ($\cos \theta_m  \ge 0.89$),
$D$ still follows an Arrhenius law, but with an activation energy of $2\epsilon$, i.e. four times
 the corresponding value for $\alpha_{bb}$.  
This functional dependence is the same as the one that quantifies the number of  particles with four broken
bonds, i.e.  particles completely detached from the network. This suggests that ---
consistent with previous  results of the slow dynamics of  models of water\cite{poolewater} and other tetrahedral networks\cite{pwm,DNA} --- 
 the diffusion of these  rare particles dominates the microscopic dynamics.   
 
Independently from the microscopic dynamics,   $D$ always vanishes  following the  Arrhenius behavior characteristic of strong atomic and molecular network forming liquids.   The activation energy value (which can vary up to a factor of four) encodes information  about  the dominant microscopic mechanism.   

  In summary, our calculations clearly show that patchy colloids with a limited number of flexible bonds
 smoothly transform from a liquid to a fully bonded network, via the progressive reduction of the number of network defects, without the intervention of a more stable crystal phase or phase separation.   Counterintuitively, the low $T$ phase behavior  is governed by entropy, rather than energy, as all competing phases reach the ground state. The  main requirements for the existence of a stable liquid at low $T$  are thus  a  large flexibility of the interparticle bonds  and a low valence. The large patch width ensures that a wide variety of network realizations can be formed. This configurational entropy is instrumental in stabilizing the liquid phase with respect to the crystal.  The low valence ensures that the  density of the liquid 
 coexisting with the gas  is  small\cite{emptyliquids}, leaving a large density  region where the network can form.  Both   are key ingredients for the formation of what can be considered a previously unexplored state of matter: the thermodynamically stable fully bonded network state.  


\section{Methods summary}

\subsection{Model}

\begin{figure}[b]
\begin{center}
\begin{tabular}{ccc}
\includegraphics[width=0.15\textwidth]{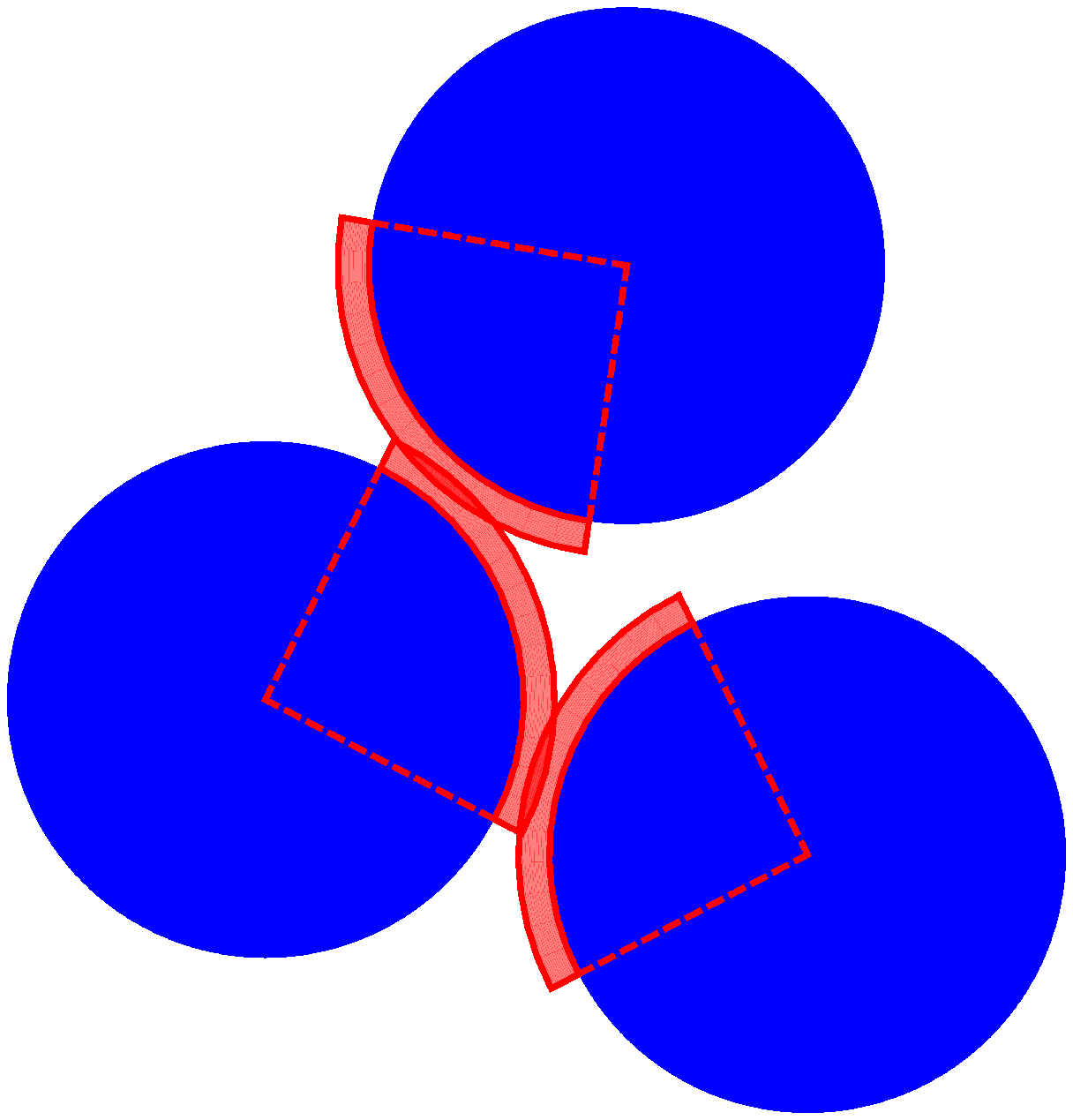} & \includegraphics[width=0.15\textwidth]{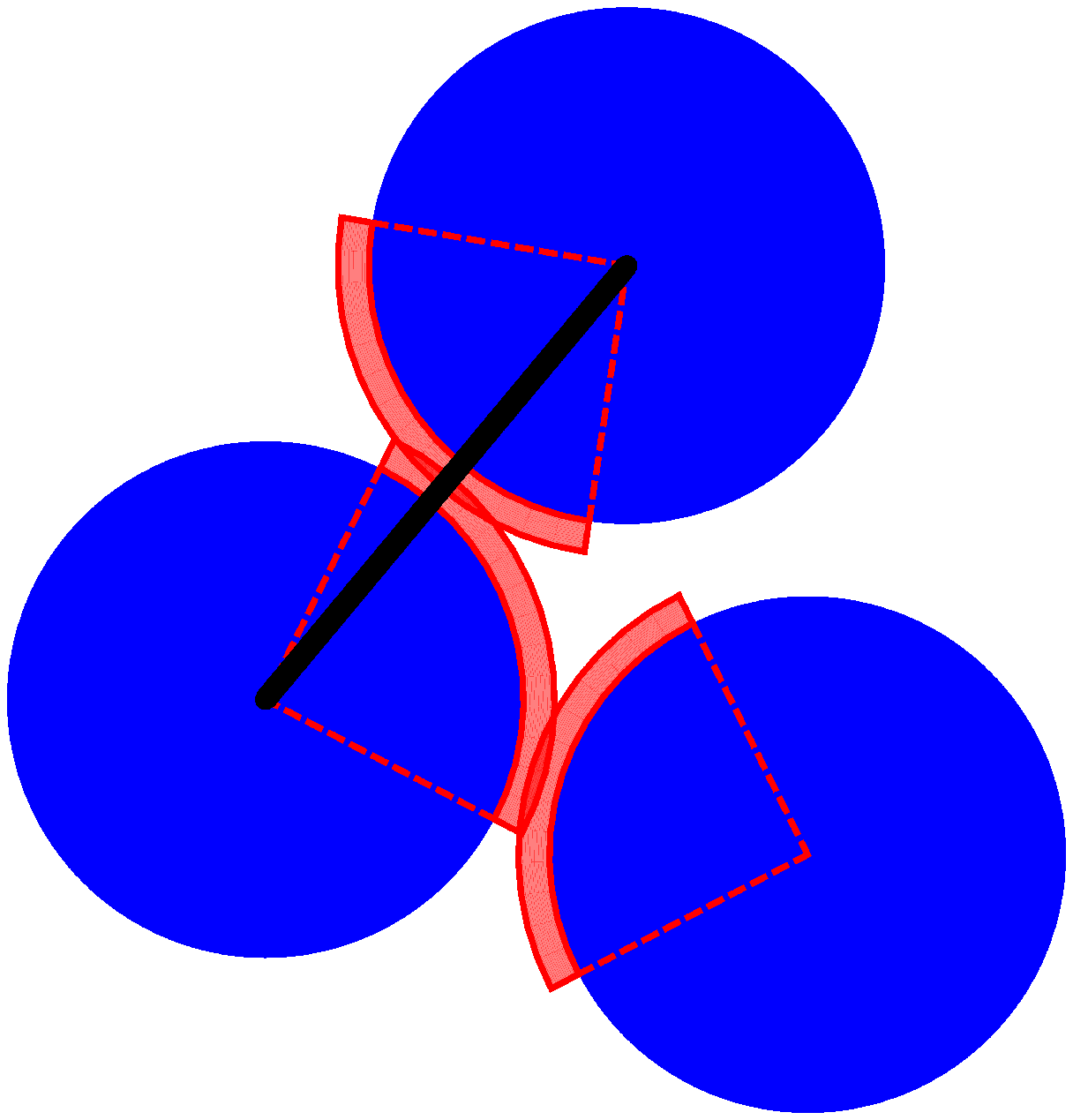} & \includegraphics[width=0.15\textwidth]{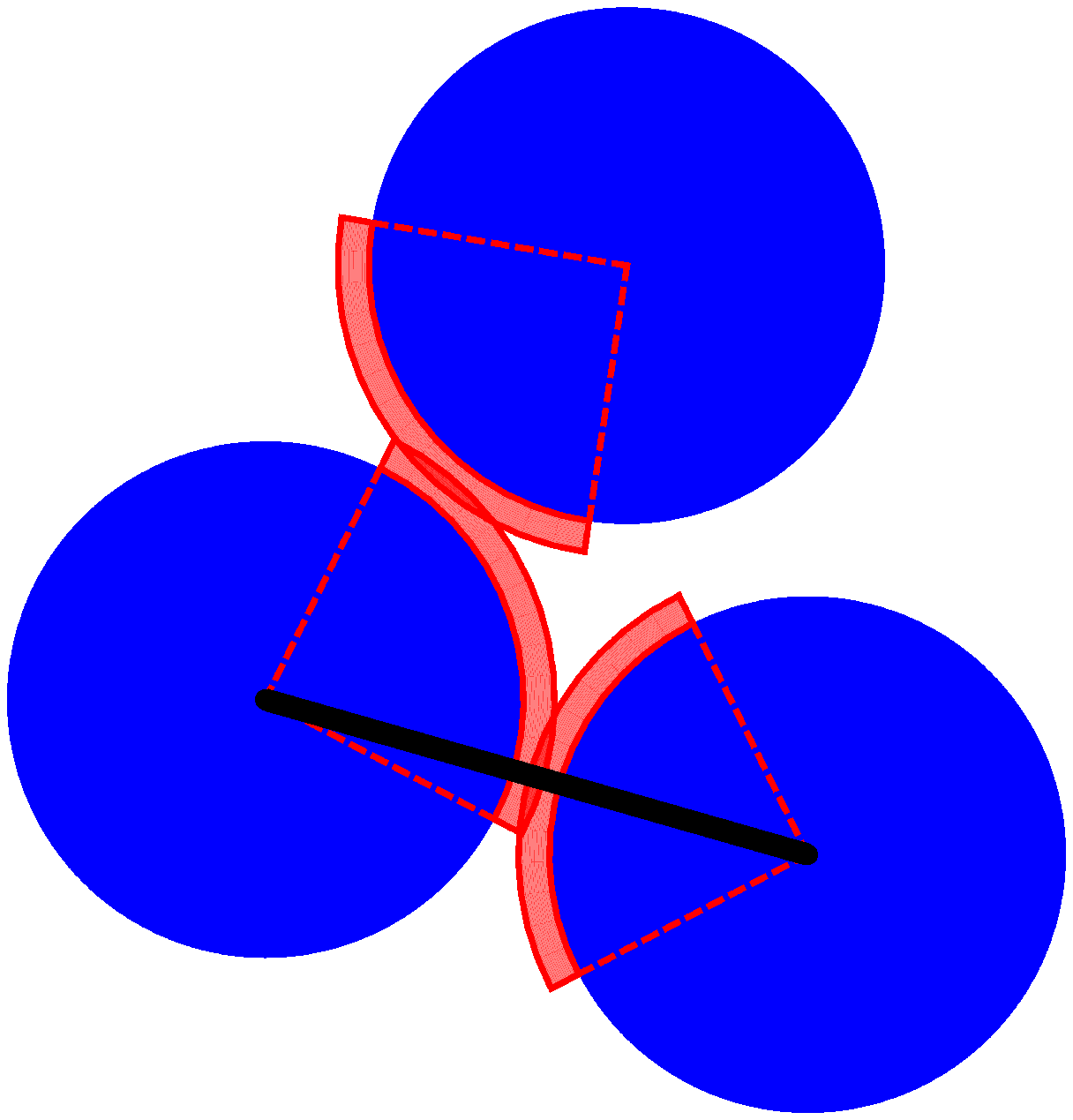}  \\
$U = 0$ & $U = -\epsilon$ & $U = -\epsilon$ 
\end{tabular}
\caption{{\bf Cartoon of the interaction between patchy particles.} In the configuration on the left, three particles are close enough together to bond, but unbonded (potential energy $U = 0$). In the next two configurations, the two different bonding possibilities between the particles are shown, both with potential energy $U = -\epsilon$. In this case, at most one bond can be formed.} \label{cartoon}
\end{center}
\end{figure}

We consider an extension of the Kern-Frenkel model\cite{KF}, where each patch can form at most one bond (see  Fig. \ref{cartoon}).
In the normal Kern-Frenkel model, two patchy particles $i$ and $j$, located at $\mathbf{r}_{i}$ and $\mathbf{r}_j$ respectively, feel an attraction given by
\begin{eqnarray}
u_{\mathrm{patch}}(i,j) =  u_\mathrm{sw}(r_{ij}) \Phi(\mathbf{r}_{ij}, \{\mathbf{p}_i\})  \Phi(\mathbf{r}_{ji},\{\mathbf{p}_j\})
\label{Utot}
\end{eqnarray}
where $\mathbf{r}_{ij} = \mathbf{r}_{j} - \mathbf{r}_{i}$,  $\{\mathbf{p}_i\}$ is set of normalized vectors pointing from 
the center of particle $i$ towards each of its patches, and $u_\mathrm{sw}$ is a square-well potential of hard diameter $\sigma$, range $\delta$ and depth 
$\epsilon$. 
\begin{equation}
\beta u_\mathrm{sw}(r) =
\left\{
	\begin{array}{ll}
		\infty & \mbox{if $r < \sigma$} \\
		-\beta\epsilon  & \mbox{if $\sigma \le r < \sigma + \delta$}\\
		\:\:\:\: \:\: 0 & \mbox{otherwise } 
	\end{array}
\right.
\end{equation}
Here, $\beta = 1/k_B T$, with $k_B$ Boltzmann's constant.

The function $\Phi(\mathbf{r}_{kl}, \{\mathbf{p}_k\})$ is defined as 
\begin{equation}
\Phi(\mathbf{r},\{\mathbf{p}\}) =
\left\{
	\begin{array}{ll}
		1  & \mbox{if $\mathbf{\hat{r}} \cdot \mathbf{\hat{p}} > \cos(\theta_{m})$ for any $\mathbf{p}$ in $\{\mathbf{p}\}$}\\
		0 & \mbox{otherwise } 
	\end{array}
\right.
\end{equation}
In short, two particles bond if the vector connecting the centers of the particles passes through an attractive patch on each of the surfaces of each particle.  Hence, 
as long as the patches on the same particle do not overlap, the Kern-Frenkel model only allows for a single bond between two particles. However, a single patch can make a bond with multiple different nearby particles if the patch width is larger  (for our $\delta$) than $0.895$.

Here, we propose and implement (see SI) a modification to the Kern-Frenkel model that allows only one bond per patch for all $\theta_m$ values.
In this modification, the overlap of the bonding volumes is not a sufficient criterium for
defining the presence of a bond.

\subsection{Methods}

To calculate the free energy of the various phases as a function of the density, we use thermodynamic integration over the equation of state\cite{bookfrenkel}, as measured using Event-driven Molecular Dynamics (EDMD) simulations.  For the gas, the reference state is an ideal gas. For the liquid, and fluids above the critical point, the hard-sphere fluid is used as a reference state, integrating over the well depth $\epsilon$ at fixed density. Similarly, the hard-sphere crystal was used as a reference state for the disordered FCC crystal. We use the Frenkel-Ladd method to calculate free energies of the diamond, BCC, and ordered FCC crystal structures\cite{frenkladd}, using an additional aligning potential to fix the orientation of the particles\cite{noya_2007}. Finally, we determine phase coexistence using a common-tangent construction, and use the resulting coexistence points to draw the phase diagram.

For the calculation of the total entropy $S_\mathrm{tot}$ at $T=0$, we use thermodynamic integration over the temperature. As the potential energy of all phases decreases exponentially at low temperatures, extrapolation to zero temperature is straightforward, and integrating over the exponential decay allows us to directly calculate the entropy at $T=0$. The common tangent constructions at zero temperature are shown in the SI.

To calculate the vibrational entropy $S_\mathrm{vib}$ in the liquid phase, we use the Frenkel-Ladd method on a fully bonded configuration taken from a simulation at low temperature. The network topology is kept fixed during the calculation. To reduce the maximum spring constant required in the integration, we first ``optimize'' the configuration in a separate MC simulation biased towards configurations where small displacements of particles do not cause overlaps or break bonds. See the SI for more information. 


We use EDMD simulations for calculating the equations of state and potential energies required for thermodynamic integration, and to investigate the dynamics of the low-temperature liquid phase\cite{rapaport,edmdanisotropic}.
For a full description of the implementation of the Kern-Frenkel model in EDMD simulations, see the SI. For all EDMD simulations used here, the simulation box is chosen to be cubic or rectangular, with periodic boundary conditions. The simulations are performed at fixed number of particles $N$, volume $V$, and temperature $T$. The temperature is regulated by means of a thermostat: at regular intervals, a random particle is given a new velocity and angular velocity, drawn from a Maxwell-Boltzmann distribution. Time is measured in units of $\tau = \sqrt{\beta m \sigma^2}$, with $m$ the mass of a particle. For the simulations in this study, we chose the moment of inertia $I = m \sigma^2$. Note that the choice of mass or moment of inertia has no effect on the equilibrium phase behavior. Bond switching moves in the EDMD simulation (see SI) are events that happen at a fixed rate $\gamma$ for each patch in the system, such that per time unit $\tau$ each patch experiences $\gamma \tau$ attempts to form, switch, or break a bond. Diffusion coefficients were obtained from the slope of the mean squared displacement as a function of time. The system sizes (ranging from $N = 512$ to $64000$) were chosen to ensure that the number of broken bonds in the system would be at least one on average, for the temperatures under consideration.


\section{Acknowledgements}
We acknowledge support from  ERC-226207-PATCHYCOLLOIDS. We thank F. Romano for useful discussions.

\end{document}